\documentclass[prd, aps, nofootinbib, preprintnumbers, showpacs, twocolumn]{revtex4}
\usepackage{amssymb}
\usepackage{amsmath}
\usepackage[T1]{fontenc}
\usepackage[english]{babel}
\usepackage{graphicx}
\usepackage[latin9]{inputenc}
\usepackage{mathpazo}
\usepackage{colordvi}

\begin{document}

\title{A darkless spacetime}
\author{Angelo Tartaglia}
\email{angelo.tartaglia@polito.it}
\author{Monica Capone}
\email{monica.capone@polito.it}
\affiliation{Dipartimento di Fisica, Politecnico di Torino, Corso Duca degli Abruzzi 24,
I-10129 Torino, Italy and INFN, sezione di Torino, Torino, Italy}
\date{\today}

\begin{abstract}
In cosmology it has become usual to introduce new entities as dark matter
and dark energy in order to explain otherwise unexplained observational
facts. Here, we propose a different approach treating spacetime as a
continuum endowed with properties similar to the ones of ordinary material
continua, such as internal viscosity and strain distributions originated by
defects in the texture. A Lagrangian modeled on the one valid for simple
dissipative phenomena in fluids is built and used for empty spacetime. The
internal \textquotedblleft viscosity\textquotedblright\ is shown to
correspond to a four-vector field. The vector field is shown to be connected
with the displacement vector field induced by a point defect in a
four-dimensional continuum. Using the known symmetry of the universe,
assuming the vector field to be divergenceless and solving the corresponding
Euler-Lagrange equation, we directly obtain inflation and a phase of
accelerated expansion of spacetime. The only parameter in the theory is the
\textquotedblleft strength\textquotedblright\ of the defect. We show that it
is possible to fix it in such a way to also quantitatively reproduce the
acceleration of the universe. We have finally verified that the addition of
ordinary matter does not change the general behaviour of the model.
\end{abstract}

\maketitle

%---------------------

\section{Introduction}

\label{sec:intro} %---------------------

When studying the universe as a whole we have to take into account a number
of observed behaviours and of physical constraints. It is well known that,
since its early years, General Relativity (GR) provided cosmological
solutions able to describe the large scale evolution of the universe from a
singular event (the Big Bang) to the present epoch. While accumulating
evidence, however, more and more details are entering the scene and the
original theoretical framework is no more enough to account for all of them.
This is the reason why people have tried and are trying to introduce new
theories of gravity, alternative to the original GR, or to modify it in a
way or another.

Since the moment when growing evidence supported the idea that the universe
is undergoing an accelerated expansion~\cite{expansion} Einstein's
\textquotedblleft big blunder\textquotedblright ~\cite{gamow}, the
cosmological constant, has seen a big revival and has been enriched with new
and sophisticated theories. One is led to think that the universe is filled
up of something exerting a negative pressure, responsible for the
acceleration. This \textquotedblleft something\textquotedblright\ has been
called in various ways, the most popular being dark energy which becomes for
instance quintessence \cite{quintessence} or phantom energy \cite{odyntsov}
according to some specific theories. The case of the accelerated expansion
is not the only one to be treated by means of some new field. The
homogeneity of the cosmic microwave background (CMB) radiation seems to
imply, very close to the big bang, a phase of extremely fast expansion and
this is accounted for by means of an \textit{ad hoc} scalar field, the
inflaton, with an even more \textit{ad hoc} potential \cite{inflaton}. We
also know about dark matter, needed to explain inhomogeneities of the CMB,
the rotation curve of galaxies and the behaviour of galaxy clusters \cite%
{darkmatter}. Yet another approach consists in using a modified
Hilbert-Einstein action integral, expressed in terms of some non-linear
function of the scalar curvature \cite{wands,higherorder}.

The situation, even though being richer and more varied, resembles the one
with ether at the end of the XIX century, and Occam's razor seems to be left
aside for a while.

Here we try a different approach taking advantage of analogies with other
branches of physics. The power of analogical deduction has played an
important role in the past and still proves to be fruitful even today, for
instance in the case of black holes and Hawking radiation \cite{analog}.

Our current vision of the cosmos, especially in GR, is essentially
dualistic, the actors being spacetime on one side (left hand side of the
Einstein's equations) and matter-energy on the other (right hand side of the
equations). Structures, differences, variety of features belong to
matter-energy. The only intrinsic property of spacetime, besides the ones
induced by matter-energy through the coupling constant $G$, is expressed by
the signature of the metric tensor.

The paradigm we are proposing here considers a spacetime endowed with some
more features on its own that remind those of a physical continuum.
Whenever, in a given physical system, we find a symmetry, we know that
something real must be there to induce that symmetry. In the case of the
whole universe, its global symmetry, in four dimensions, implies the
presence of a singular event: the center of symmetry. We may state it either
way: telling that the symmetry implies a zero-dimensional singularity, or
that the singularity induces the symmetry. The novelty in our approach is in
thinking that the singularity is not due to the content (mass-energy) of the
spacetime, but is built in the very spacetime. The next step consists in
interpreting and treating the singularity just as a defect in a continuous
medium is in the classical elasticity theory \cite{eshelby,Kleinert}. A
point defect in an otherwise homogeneous and isotropic medium induces a
strained state, characterizable by means of a radial vector field: the
\textquotedblleft rate of stretching\textquotedblright\ in the radial
position. That vector field is divergence-free, except in the center of
symmetry.

In our theory it is indeed a four-vector field that plays an important role
and it is not at all the first time that a vector field is introduced in the
description of the behaviour of the cosmos. An example is Bekenstein
modified theory of gravity (a proposed relativistic version of Milgrom's
MOND \cite{milgrom}) which can be thought of as a scalar-vector-tensor
theory \cite{zlosnik}. Besides this, the core of the so called
Einstein-aether theory (Æ-theory) is a timelike unit vector field \cite%
{eling} which is introduced in the spacetime action integral in the way this
is usually done for additional field components. In the Einstein-aether
theory, in particular, the cosmic vector field appears in the Lagrangian
through its first covariant derivatives and a Lagrangian multiplier related
to the unit norm constraint. The vector field brings about a violation of
the Lorentz symmetry, a priori\ introducing a preferred local rest frame. Æ%
-theory has been explored in its consequences (see for instance \cite%
{brendan}, \cite{lim}, and the references quoted in \cite{eling}) in order
to fix upper bounds on its parameters and concluding that it would be
compatible with the present state of observations and experiments \cite%
{brendan}. Our approach is however different; for us, the primordial fact is
a pointlike defect in spacetime and it is this defect that induces both the
symmetry of the physical manifold we call spacetime, and the vector field.
In this way the form of the field is determined by the very existence of the
defect, just in the same way as a strain state is induced by defects in
material continua.

The idea of spacetime as an elastic continuum with properties depending on
its \textquotedblleft microscopic structure\textquotedblright\ has a story
on its own, with illustrious antecedents, such as Sakharov \cite{sakharov},
who tried to explain the \textquotedblleft rigidity\textquotedblright\ and
Lorentz invariance of the \textquotedblleft medium\textquotedblright\ in
terms of zero point fluctuations of quantum fields. Presently, Loop Quantum
Gravity theories \cite{rovelli} consider a sort of atomistic description of
spacetime. On the other side, the theory of defects is a well established
one and was developed more than one century ago together with the elasticity
theory; the idea of extending it to more than three dimensions is not new
\cite{Kleinert,katanaev,malyshev,difetti,padmana}, but it has never been
considered more than a curiosity. Recently, the idea of some
\textquotedblleft solid behaviour\textquotedblright\ has been called in for
dark energy \cite{battye}; there, however, the solid is the dark energy
itself and its description is the one of a three-dimensional medium evolving
in time. Our continuum instead is the very spacetime and its solid
properties are described in four dimensions. As it is the case for classical
GR, we study a global equilibrium state (including strain, distortion and
whatever else) from the center of symmetry (the defect) to infinity.

In order to write down the appropriate action integral for the spacetime
considered above (including the defect) we remark that the phase space of
the system is bidimensional, the generalized coordinates being the scale
factor and its rate of change. A similar phase space, whose generalized
coordinates are position and velocity, is the one describing the motion of a
massive point particle across a viscous fluid. From this starting point, we
are able to write down and then generalize an appropriate Lagrangian. What
we obtain in the end is a spacetime displaying inflationary expansion in the
neighbourhood of the center of symmetry (i.e. the Big Bang), then a
deceleration-acceleration-deceleration sequence.

The theory does not exclude the actual presence of matter-energy; in order
to study its effect on the behaviour of the universe as a whole, we
introduce it in the form of an ordinary perfect fluid, as usual. We find
that in the negligible pressure era (today) the presence of matter has no
influence on the global solution. In the radiation dominated era the general
behaviour is preserved when the matter-energy content is lower than a
critical value. The theory contains one free parameter, which is the
``size'' or ``strength'' or ``charge'' of the singularity. We may fix it in
such a way that the present value of the Hubble constant as well as the age
of the universe are reproduced. Doing so, we see that the currently
estimated content of matter-energy in the cosmos is well below the critical
value, and the position and duration of the accelerated expansion are
consistent with the data from observations.

The paper is organized as follows. In Sec.~\ref{sec:class} we study, from
the viewpoint of variational methods, the simple classical problem of a
particle moving in a dissipative medium and extend it to the relativistic
formalism. In Sec. ~\ref{sec:spacetime} a \textquotedblleft
dissipative\textquotedblright\ Lagrangian for spacetime is introduced, and
then specialized to the case of a homogeneous and isotropic empty universe;
in Sec. ~\ref{sec:matter} we analyze the effect of the inclusion of ordinary
matter; Sec. \ref{sec:Newton} verifies the existence of the Newtonian limit
of the theory. Finally Sec. ~\ref{sec:conclusions} is devoted to the summary
of our findings and to the discussion of our conclusions. The signature used
in the paper for the metric tensor is ($+,-,-,-$).

%----------------------------------------------------------

\section{A classical model: a particle in a viscous medium}

\label{sec:class}
%----------------------------------------------------------

\begin{figure}[t]
\begin{center}
\includegraphics[width=80 mm, height=71 mm]{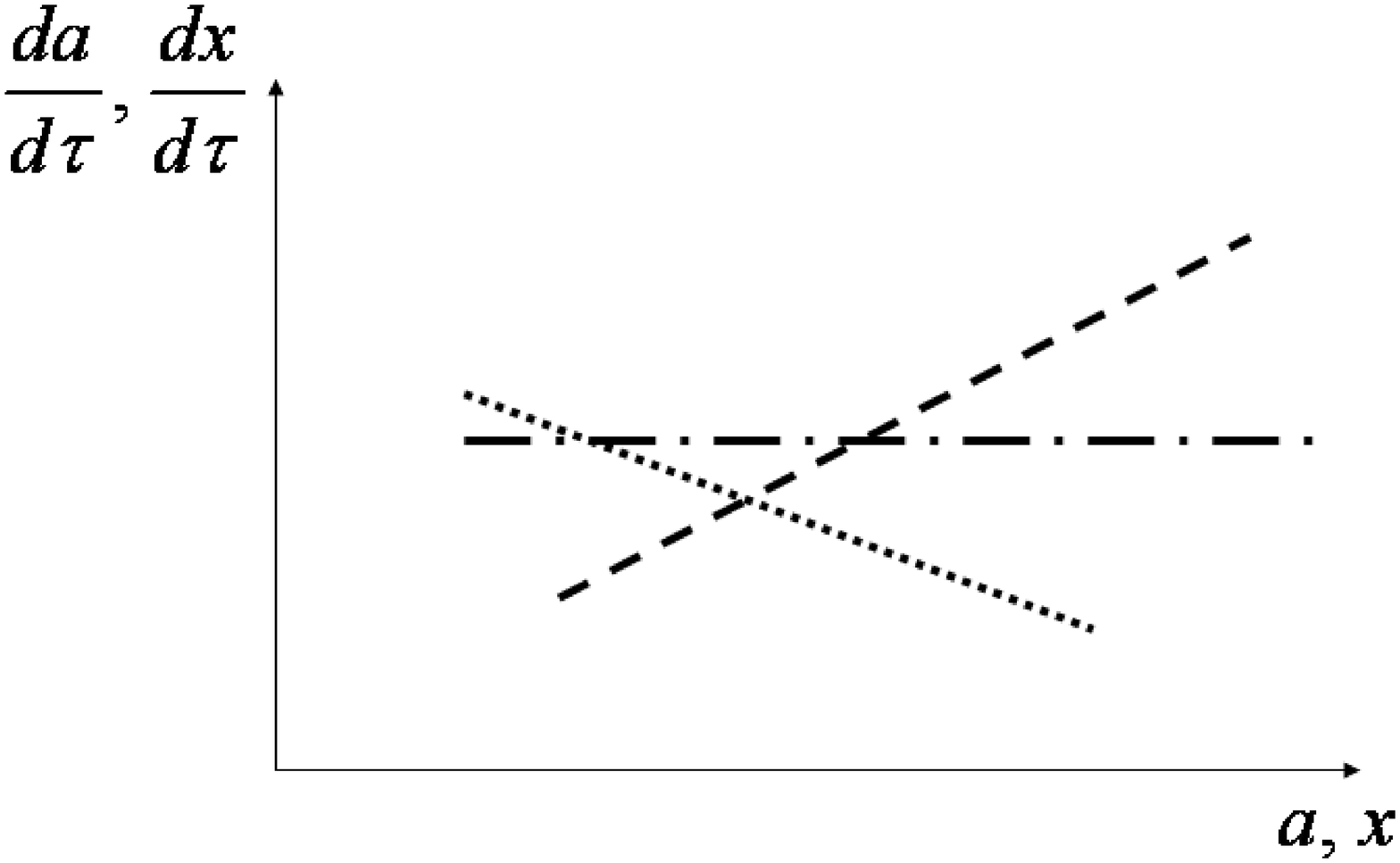}
\caption{\label{fig:fig0}Phase diagram of a Robertson Walker
universe. The dash and dot line corresponds to an inertial
expansion (constant expansion rate); the dotted line is a
decelerated expansion; the dashed line is an accelerated
expansion. Observations tell us that the actual behaviour of the
universe corresponds to both deceleration and acceleration, in
different epochs. Relabelling the axes with an $x$ instead of $a$,
and a $dx/dt$ instead of $da/d\protect\tau $, the diagramme
describes the state of a point particle interacting with a
surrounding medium. Dash and dot is inertial motion (constant
speed); dots is motion under the action of friction; dash is an
accelerated motion driven by the medium.}
\end{center}
\end{figure}

Once one has decided to try and account for a given observation
(in our case it is the accelerated expansion of the universe) by
modifying a consolidated theory, such as GR, one needs some
criterion to decide what change to introduce and test. A possible
and often used approach is to explore mathematical variants of the
basic theory, introducing free functions and free parameters, then
adjusting the parameters and the functions so to reproduce the
observed results and satisfy the physical constraints the problem
has. This procedure can be more or less fortunate, but often poses
problems of physical interpretation with the newly introduced
functions and parameters. A different way may be, as already
mentioned in the introduction, to look for correspondences with
other parts of physics. In fact we know that in many cases the
same set of equations can govern apparently unrelated phenomena.
This is the case for instance of classical field theory and
fluidodynamics, elastic waves in solids and fluids and
electromagnetic waves, conservation laws in general, etc. For this
very reason we decided to look around for some ordinary situation
displaying the same formal properties as a Robertson Walker (RW)
universe. Indeed, if we accept that the universe has the typical
symmetry expressed by the RW line element, we see that the state
of the universe is described by the only scale factor $a$ with an
evolution parameter $\tau $ (cosmic time). The situation may be
schematized as in fig. (\ref{fig:fig0}), where various evolution
trends are drawn. A simple transliteration (from $a$ to $x$)
converts the evolution of the universe into the interaction of a
point particle with an isotropic fluid.

%
%%%%%%%%%%%%%%%%%%%%%%%%%%%%%%%%%%%%%%%%%%%%%%%%%%%%%%%%%%%%%%%%%%%%%%%%%%%%
%           FIGURE 0
%%%%%%%%%%%%%%%%%%%%%%%%%%%%%%%%%%%%%%%%%%%%%%%%%%%%%%%%%%%%%%%%%%%%%%%%%%%%
%%%%%%%%%%%%%%%%%%%%%%%%%%%%%%%%%%%%%%%%%%%%%%%%%%%%%%%%%%%%%%%%%%%%%%%%%%%%%

Exactly the same type of phase space is obtained when describing the motion
of a point particle interacting with a surrounding fluid. Now the
interesting feature of this analogy is that we know how to write the
equations governing the motion of a point particle of mass $m$ in a viscous
fluid. Although dissipative, the problem can be treated starting from the
Lagrangian
\begin{equation}
L=e^{\gamma t+\eta x}\frac{m}{2}\dot{x}^{2}\ .  \label{lag}
\end{equation}%
A simpler form of this expression was initially introduced by Caldirola~\cite%
{cald}, then by others~\cite{oth} for different purposes. The Euler-Lagrange
equation deduced from Eq.~(\ref{lag}) is
\begin{equation}
\ddot{x}+\gamma \dot{x}+\dfrac{\eta }{2}\dot{x}^{2}=0\ ,  \label{moto}
\end{equation}%
where $\gamma >0$ may be interpreted as the laminar viscosity coefficient
and $\eta $ is the turbulent viscosity coefficient. Actually, Eq.~(\ref{moto}%
) represents a viscous motion if for $\eta >0$ the motion is progressive;
for regressive motion $\eta $ has to be assumed $<0$. In this approach, the
properties of the fluid and of the interaction are all contained in $\gamma $
and $\eta $, that are assumed to be constants, which means that the fluid is
not affected by the motion of the particle through it. Eq.~(\ref{moto}) is
however inadequate since it is not invariant for reversal of the $x$-axis.
If $v_{0}$ is the initial velocity of the particle, the solution of (\ref%
{moto}) is
\begin{equation}
\dot{x}=\frac{2\gamma }{\left( \dfrac{2\gamma }{v_{0}}+\eta \right)
e^{\gamma t}-\eta }\ ,
\end{equation}%
For $-\infty <v_{0}<-\frac{2\gamma }{\eta }$, the solution diverges at some
finite positive time. We now recast the problem in a relativistic way. We
shall consider a flat spacetime and introduce the action integral
\begin{equation}
S=-m\int_{A}^{B}e^{\alpha \tau ^{\prime }-\beta x^{\prime }}ds\ ,
\label{action0}
\end{equation}%
where $d\tau ^{\prime }=cdt^{\prime }$, $v^{\prime }=\dot{x}^{\prime }$ and $%
ds=\sqrt{1-v^{\prime 2}/c^{2}}d\tau ^{\prime }$. The exponent in Eq.(\ref%
{action0}) is assumed to be a true scalar, i.e. the scalar product of two
four-vectors
\begin{eqnarray*}
\gamma &=&\left( \alpha ,\beta ,\beta ,\beta \right) \ , \\
r^{\prime } &=&\left( ct^{\prime },x^{\prime },0,0\right) \ .
\end{eqnarray*}%
The reference frame has been fixed so that the $x^{\prime }$ axis coincides
with the direction of motion. Using Cartesian coordinates, the form of $%
\gamma $ expresses the expected space isotropy of the medium. The invariant
associated with $\gamma $ is
\begin{equation*}
\chi ^{2}=\alpha ^{2}-3\beta ^{2}>0\ ,
\end{equation*}%
and the four-vector has been assumed to be timelike. The Lorentz-invariant
form of Eq.~(\ref{action0}) is now
\begin{equation}
S=-m\int_{A}^{B}e^{\eta _{\alpha \beta }\gamma ^{\alpha }x^{\beta }}ds\ ,
\label{action1}
\end{equation}%
The Euler-Lagrange equation from (\ref{action0}) is
\begin{equation}
\ddot{x}^{\prime }-\frac{\beta }{2}c^{2}\left( 1-\frac{\dot{x}^{\prime 2}}{%
c^{2}}\right) ^{2}+\frac{\alpha }{c}\left( 1-\frac{\dot{x}^{\prime 2}}{c^{2}}%
\right) \dot{x}^{\prime }=0\ .
\end{equation}

Everything becomes more transparent and simpler if, applying an appropriate
Lorentz transformation, we change the reference frame so that
\begin{equation}
\gamma =(\chi ,0,0,0)\ .  \label{vector}
\end{equation}%
We see that in this case a privileged reference frame exists: it is the one
of the fluid (unprimed quantities). The equation of motion is now
\begin{equation}
\ddot{x}+\chi \left( 1-\frac{\dot{x}^{2}}{c^{2}}\right) \frac{\dot{x}}{c}=0\
.  \label{motion}
\end{equation}%
Equation(\ref{motion}) represents the relativistic version of motion in
presence of laminar viscosity; now the solution is a decelerated motion and
no troubles arise from any reversal of the space axes. It is explicitly:
\begin{equation}
\dot{x}=\pm \dfrac{v_{0}}{\sqrt{\dfrac{v_{0}^{2}}{c^{2}}+\left( 1-\dfrac{%
v_{0}^{2}}{c^{2}}\right) e^{\frac{2\chi t}{c}}}}\ ,
\end{equation}
Starting with an initial value $v_{0}<c,$ the velocity becomes zero in an
infinite time; \textquotedblleft photons\textquotedblright\ do not interact
with the medium (their velocity stays equal to $c$). We could reasonably
introduce a dependence of $\chi $ on $v$, but to discuss further this
elementary situation is out of the scope of the present paper. What matters
is that it is possible to give a Lagrangian treatment of a simple
dissipative phenomenon describing a non-uniform evolution in time, in a
relativistic context.

%---------------------------------------

\section{The behaviour of spacetime}

\label{sec:spacetime} %---------------------------------------

We may now use the model in the previous section as a guiding idea (by
analogy) and the action~(\ref{action1}) as a suggestion or inspiration to
describe a spacetime that undergoes expansion or contraction at a
non-uniform rate. We stress that not the whole universe but the only
spacetime is considered.

The starting point is the usual Einstein-Hilbert action
\begin{equation}  \label{EH}
S=\int_{\Omega _{1}}^{\Omega _{2}}Rd\Omega \ ,
\end{equation}
where $R$ is the scalar curvature and $d\Omega =\sqrt{\left\vert
g\right\vert }d^{4}x$ is the invariant volume element.

We directly introduce in the spacetime the kind of symmetry we usually
attribute to the universe, i.e. four-dimensional isotropy around a given
event. As it is well known the most general symmetric line element of this
type is the RW one
\begin{equation}
ds^{2}=d\tau ^{2}-a\left( \tau \right) ^{2}\left[ \frac{dr^{2}}{1-kr^{2}}%
+r^{2}\left( d\theta ^{2}+\sin ^{2}\theta d\phi ^{2}\right) \right] \ ,
\label{RW}
\end{equation}%
where $k=0$,$\pm 1$ and $r$ is a dimensionless coordinate (product of usual
length times the square root of the space curvature). Introducing the metric
tensor implicit in Eq.~(\ref{RW}) into Eq.~(\ref{EH}) one has
\begin{equation}
S=-6\mathcal{V}_{k}\int_{\tau _{1}}^{\tau _{2}}\left( a\ddot{a}+\dot{a}%
^{2}+k\right) ad\tau \ ,  \label{action2}
\end{equation}%
where $\mathcal{V}_{k}=\int_{0}^{r}\int_{0}^{\pi }\int_{0}^{2\pi }\frac{%
r^{2}\sin \theta }{\sqrt{1-kr^{2}}}drd\theta d\phi $ and dots represent
derivatives with respect to $\tau $.

%----------------------------------------

\subsection{``Dissipative'' spacetime}

\label{sec:diss_spacetime} %----------------------------------------

To take advantage of the simple example described in Sec.~\ref{sec:class},
we may want to reproduce the same logical structure while building the
action integral. Of course, there are important differences to take into
account. In Sec.~\ref{sec:class} we had two \textquotedblleft
actors\textquotedblright\ entering the scene, the particle and the
dissipative medium, and an interaction between them. The interaction was
mediated by a four-vector pertaining to the medium and another one
describing the state (of motion) of the particle. Now the \textquotedblleft
actor\textquotedblright\ is unique, spacetime itself, and nothing is moving
across it. However, we may think to the motion of the representative point
of the state of a hypersurface labelled by $\dot{a}$ in a bidimensional
phase space where the independent variable is the parameter $a$. Again, we
associate to the system a four-vector $\gamma $ which couples to the state
variable in the same way as in the simple expression~(\ref{action1}). There
the position coordinate of the particle appeared in the Lagrangian through
its (first) derivative $dx/dt$; now, the Lagrangian is written in terms of
the (second) derivatives of the elements of the metric tensor. The simplest
way to couple a vector to the metric in order to generate a scalar is by
taking the norm of the vector itself, so we are led to conjecture the
following action integral
\begin{equation}
S=\int_{\Omega _{1}}^{\Omega _{2}}e^{\pm g_{\mu \nu }\gamma ^{\mu }\gamma
^{\nu }}Rd\Omega \ ,  \label{action3}
\end{equation}%
where $\gamma $ is meant to represent an \textit{internal property} of
spacetime. The correspondence between (\ref{action3}) and (\ref{action1})
may be better seen considering that in (\ref{action1}) the position vector
of the particle is made out of the coordinates of the particle, whose choice
is free, and the final behaviour of the system is independent from that
choice. In the case of (\ref{action3}), i.e. for spacetime, the
"coordinates" used to describe the state of the system are represented by
the elements of the metric tensor and again there is a gauge freedom in
their choice. As written above, the simplest scalar we can build from a
vector and the metric tensor is the one in the exponent of (\ref{action3}).

Using the metric in the form~(\ref{RW}), in the (privileged) cosmic
reference frame, and the four-dimensional rotation symmetry, $\gamma $
necessarily appears in the \textquotedblleft radial\textquotedblright\ form
of Eq.~(\ref{vector}), so that the action~(\ref{action3}) reads
\begin{equation}
S=-6\mathcal{V}_{k}\int_{\tau _{_{1}}}^{\tau _{_{2}}}e^{\pm \chi ^{2}}\left(
a\ddot{a}+\dot{a}^{2}+k\right) ad\tau \ .  \label{action5}
\end{equation}%
The $\chi $ = constant case is irrelevant because it corresponds to simple
Minkowski spacetime (the exponential may be absorbed in a rescaling of the
time coordinate). Non-trivial solutions exist only if
\begin{equation}
\chi \equiv \chi \left( a\left( \tau \right) \right) \ ,
\end{equation}%
and spacetime is not globally homogeneous and isotropic. The observation of
the universe apparently tells us that space is flat, so we further assume $%
k=0$. The final effective Lagrangian is
\begin{equation}
\mathcal{L=}e^{\pm \chi ^{2}}\left( a\ddot{a}+\dot{a}^{2}\right) a\ .
\label{lagra}
\end{equation}%
The Euler-Lagrange equation for $a\left( \tau \right) $ is
\begin{equation}
\frac{d^{2}}{d\tau ^{2}}\left( \frac{\partial \mathcal{L}}{\partial \ddot{a}}%
\right) -\frac{d}{d\tau }\left( \frac{\partial \mathcal{L}}{\partial \dot{a}}%
\right) +\frac{\partial \mathcal{L}}{\partial a}=0\ ,
\end{equation}%
or explicitly
\begin{align}
& \ddot{a}\left( 1 \pm 2\chi \chi ^{\prime }a\right)  \label{base} \\
& +\dfrac{\dot{a}^{2}}{a}\left\{ a^{2}\left( 2\chi ^{2}\chi
^{\prime 2}\pm \chi ^{\prime 2}\pm \chi \chi ^{\prime \prime
}\right) \pm 3a\chi \chi ^{\prime }+\dfrac{1}{2}\right\} =0\ .
\notag
\end{align}%
where $\chi ^{\prime }\equiv d\chi /da$. A trivial and non interesting
solution is obtained for $a=\mathrm{constant}$ (Minkowski spacetime). Other
solutions however exist. The above equation may be reorganized in the form
\begin{equation}
\dfrac{\ddot{a}}{\dot{a}}=-f(a)\dot{a}\ ,  \label{intermedio}
\end{equation}%
where
\begin{equation}
f(a)=\dfrac{2a^{2}\left( 2\chi ^{2}\chi ^{\prime 2}\pm \chi
^{\prime 2}\pm \chi \chi ^{\prime \prime }\right) \pm 6a\chi \chi
^{\prime }+1}{2a\left( 1\pm 2a\chi \chi ^{\prime }\right) }\ .
\end{equation}%
A first integration leads to
\begin{equation}
\dot{a}=Ae^{-\int^{a}f\left( \zeta \right) d\zeta }\ ,  \label{int1}
\end{equation}%
and finally to
\begin{equation}
\tau =\frac{1}{A}\int_{0}^{a}e^{\int^{\varsigma }f\left( \zeta
\right) d\zeta }d\varsigma \ .
\end{equation}

%----------------------------------------------------

\subsection{Choosing $\protect\chi \left( a\right) $}

\label{sbsc:chi} %----------------------------------------------------

The function $\chi(a)$ in our case is something which is ``given'' exactly
as the global symmetry is. We shall discuss this issue in more detail in the
next section. What matters here is that $\chi(a)$ is not deducible from a
variational principle. Should we try and write its field equation from the
action of Eq.~(\ref{action5}), the result would trivially be $\chi=0$, i.e.
empty and flat spacetime. A singularity, as well as a symmetry, is here an
initial condition and not a consequence of something else. This means that
we need criteria and guesses to think about credible forms for $\chi(a)$,
consistent with the hypotheses. One reasonably simple criterion for a vector
field that is expected not to spoil the symmetry we assumed for spacetime,
is to constrain it to be divergence free everywhere except in the center of
symmetry/origin of the cosmic times. In fact, any event where the divergence
of the vector differs from zero may be thought of as a ``source'' and, since
we need to preserve homogeneity and isotropy, we would have to assume a
continuous and uniformly distributed source. Being, in this paradigm, $%
\gamma $ a concrete quantity strictly related to the presence of a
singularity, it seems more reasonable to have just one pointlike source in
the origin. There is however also a different, though somehow equivalent,
reason for choosing $\gamma $ to be divergenceless, as we shall see further
on (Sec.~\ref{sbsc:defects} ), when considering the analogy with a defected
solid.

The null--divergence condition is formally written as
\begin{equation}
0=\gamma _{\;\;;\mu }^{\mu }=\left( \sqrt{\left\vert g\right\vert }\gamma
^{\mu }\right) _{,\mu }\ ,  \label{div0}
\end{equation}%
where the semicolon represents a covariant derivative. The solution of this
equation is
\begin{equation}
\chi =\frac{Q^{3}}{a^{3}}\ .  \label{csidia}
\end{equation}%
In practice, the vector field looks like the \textquotedblleft
electric\textquotedblright\ field of a point charge $Q^{3}$, but in four
dimensions. Inserting this result in Eq.~(\ref{intermedio}), and expressing $%
a$ in units of $Q$, we obtain
\begin{equation}
\frac{\ddot{a}}{\dot{a}}=-\dfrac{36\pm 24a^{6}+a^{12}}{2a^{7}(a^{6}\mp 6)}%
\dot{a}\ .  \label{a2dot}
\end{equation}%
As a consequence, we have
\begin{equation}
\dot{a}=\sqrt{\dfrac{a^{5}}{a^{6}\mp 6}\;\,e^{\mp 1/a^{6}}}\ .  \label{adot}
\end{equation}%
Choosing the upper signs in (\ref{adot}), as well as in (\ref{action5}), the
expansion rate acquires real values starting from a finite non-zero value of
$a$, then it displays a monotonic trend which does not correspond to
observational data. Choosing the lower signs instead, we see that the
expansion rate $\dot{a}$ has two extrema at
\begin{equation}
a_{M\pm }=\sqrt[3]{3\pm \sqrt{3}}\ .  \label{estremi}
\end{equation}%
The corresponding explicit numeric values for the scale factors are%
\begin{eqnarray}
a_{M_{-}} &=&1.08\ ,  \label{flexi} \\
a_{M_{+}} &=&1.68\ .  \label{flex}
\end{eqnarray}%
%
%
%
%
%
%
%
%
%
%
%
%%%%%%%%%%%%%%%%%%%%%%%%%%%%%%%%%%%%%%%%%%%%%%%%%%%%%%%%%%%%%%%%%%%%%%%%%
%                         FIGURE 1: da/dt
%%%%%%%%%%%%%%%%%%%%%%%%%%%%%%%%%%%%%%%%%%%%%%%%%%%%%%%%%%%%%%%%%%%%%%%%%
\begin{figure}[t]
\begin{center}
\includegraphics[width=90 mm, height=80 mm]{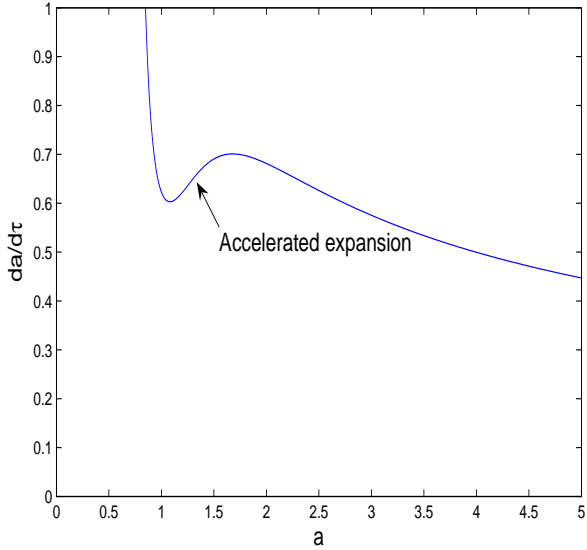}
\caption{\label{fig:fig1}Plot of $da/d\protect\tau $ as a function
of $a$, in empty spacetime. Two extrema delimiting a phase of
accelerated expansion are clearly seen.}
\end{center}
\end{figure}
%%%%%%%%%%%%%%%%%%%%%%%%%%%%%%%%%%%%%%%%%%%%%%%%%%%%%%%%%%%%%%%%%%%%%%%%%%

\begin{figure}[t]
\begin{center}
\includegraphics[width=90 mm, height=80 mm]{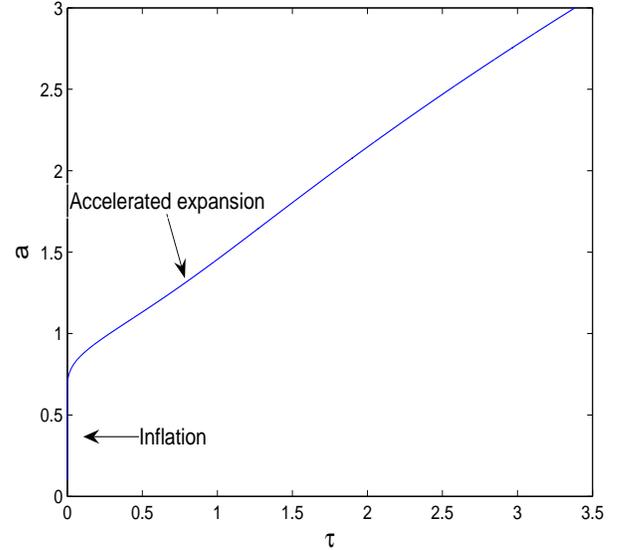}
\caption{\label{fig:adita}Scale factor $a$ as a function of cosmic
time $\protect\tau $; arbitrary units. The initial inflationary
epoch as well as an accelerated expansion between $\protect\tau
\simeq 0.41$ and $\protect\tau \simeq 1.3$ are clearly visible.}
\end{center}
\end{figure}

Fig.~\ref{fig:fig1} shows the behaviour of $da/d\tau $ as a
function of the scale factor $a$. In any case, the asymptotic
behaviour when $a\rightarrow \infty $ is $\dot{a}\rightarrow 0$:
this is a never-ending expansion. From now on, we limit our
consideration to the latter choice of signs, so, integrating
Eq.~(\ref{adot}) one has
\begin{equation}
\tau =\int_{0}^{a}\frac{\left( \zeta ^{6}+6\right) ^{1/2}}{\zeta ^{5/2}}%
e^{-1/2\zeta ^{6}}d\zeta \ .  \label{flessi}
\end{equation}%
The corresponding behaviour of the scale parameter $a$ as a function of the
cosmic time $\tau $ is shown in Fig.~\ref{fig:adita}. %
%%%%%%%%%%%%%%%%%%%%%%%%%%%%%%%%%%%%%%%%%%%%%%%%%%%%%%%%%%%%%%%%%%%%%%%%%
%                         FIGURE 2: da/dt
%%%%%%%%%%%%%%%%%%%%%%%%%%%%%%%%%%%%%%%%%%%%%%%%%%%%%%%%%%%%%%%%%%%%%%%%%
%%%%%%%%%%%%%%%%%%%%%%%%%%%%%%%%%%%%%%%%%%%%%%%%%%%%%%%%%%%%%%%%%%%%%%%%%%
%
Close to the origin the negative exponential factor in the integral in (\ref%
{flessi}) brings about an inflationary phase, that cannot be
resolved in Fig.~\ref{fig:adita}, followed by a
deceleration-acceleration-deceleration sequence driving an
unlimited expansion. In the neighborhood of the origin the scale
factor $a$ does indeed grow faster than any power of $\tau $.

We can now fix the scale of the expansion. We know that in general
\begin{equation}  \label{red}
\frac{a_{0}}{a}=1+z \ ,
\end{equation}
where $a_{0}$ is the present value of the scale factor and $z$ is the
redshift of light emitted when the scale factor was $a$. From the numerical
values~(\ref{flexi}) and~(\ref{flex}) we see that the ratio between the
scale factor at the end and at the beginning of the acceleration epoch is
\begin{equation}
\frac{a_{M+}}{a_{M-}}=1.55 \ .
\end{equation}
Then, using (\ref{red}), we can fix
\begin{equation}  \label{risultato}
\frac{1+z_{i}}{1+z_{f}}=1.55 \ ,
\end{equation}
where now $z_i$ corresponds to the redshift at $a_{M-}$ and $z_f$ to that at
$a_{M+}$. Looking at the data from the observation of high redshift Ia
supernovae ~\cite{perlmutter}, we see that (\ref{risultato}) is indeed
consistent with an initial value of $z_{i}\sim $ $1.6$ (or a little more)
and a final one $z_{f}\sim 0.6$. This result is obtained independently from
the value of the integration constant $Q$ and in the absence of matter.

An estimate of $Q$ can be obtained again from (\ref{red}). If we let the $%
a_{M-}$ value correspond to $z_{i}=1.6$, we get immediately, from (\ref{red}%
) and (\ref{flexi}), $a_{0}=2.81$, then numerically from (\ref{flessi}) $%
\tau _{0}=3.06$. Now, if $T$ is the age of the universe, it is%
\begin{equation}
\tau _{0}=\frac{cT}{Q} \ .
\end{equation}
If we constrain $T$ to be not less than 12 billion years (age of the oldest
globular cluster stars), we conclude that%
\begin{equation}  \label{stima}
Q\gtrsim 4\times 10^{25}\quad \mathrm{m} \ .
\end{equation}
This result should also be confirmed considering the present value of the
Hubble constant $H_{0}$. From (\ref{adot}) (lower signs) we see that%
\begin{equation}
\tilde{H}_{0}=\left. \frac{\dot{a}}{a}\right\vert _{\tau =\tau _{0}}= \frac{%
a_{0}^{3/2}}{\left( a_{0}^{6}+6\right) ^{1/2}}e^{1/(2a_{0}^{6})}=0.21 \ .
\end{equation}
The above result is adimensional. Introducing the appropriate dimensions it
must be%
\begin{equation}
H_{0}=\tilde{H}_{0}\frac{c}{Q}=1.6\times 10^{-18}\text{ s}^{-1}
\end{equation}%
or, using the commonly used units,
\begin{equation}
H_{0}\simeq 49\text{ km/s}\times \text{Mpc,}
\end{equation}%
Considering the roughness of the model, this is a reasonable number for the
Hubble constant, which is currently estimated to be $65$ km/s$\times $Mpc ~%
\cite{conley}.

We would like to stress that all this comes from an internal property of
spacetime, \textit{with no matter inside. }Matter must be further added to
the Lagrangian in the traditional (additive) way and with a minimal coupling
to spacetime via the metric tensor.

%--------------------------------------------------

\subsection{What could the vector field represent?}

\label{sbsc:defects} %--------------------------------------------------
%%%%%%%%%%%%%%%%%%%%%%%%%%%%%%%%%%%%%%%%%%%%%%%%%%%%%%%%%%%%%%%%%%%%%%%%%
%                   FIG. 3: the punctual defect
%%%%%%%%%%%%%%%%%%%%%%%%%%%%%%%%%%%%%%%%%%%%%%%%%%%%%%%%%%%%%%%%%%%%%%%%%
\begin{figure}[t]
\begin{center}
\includegraphics[width=70 mm, height=65 mm]{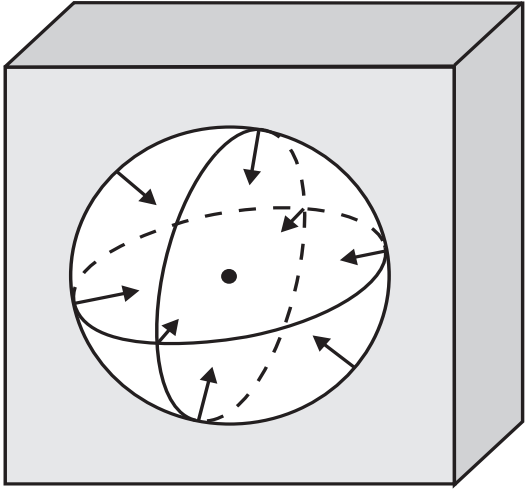}
\caption{\label{fig:defec}Point defect originated removing a solid
hypersphere, then squeezing the hole to a point.}
\end{center}
\end{figure}
%%%%%%%%%%%%%%%%%%%%%%%%%%%%%%%%%%%%%%%%%%%%%%%%%%%%%%%%%%%%%%%%%%%%%%%%%%

As mentioned above, the "internal" vector field associated with empty
spacetime can be interpreted by means of another analogy with ordinary
physics. We know that the intrinsic metric of a material continuum can be
non-Euclidean (non-zero intrinsic curvature) when defects are present (see
for example ~\cite{eshelby} or ~\cite{kat} and references therein). The
corresponding theory has been developed many years ago, starting with the
formal definition of a defect given by V. Volterra~\cite{volterra}. The
attempt to extend the theory from material elastic media to spacetime has
been made by many a scientist~\cite%
{Kleinert,atarta,katanaev,malyshev,difetti,padmana} in various epochs,
without leading to a complete formal new theory. The similarities are indeed
tempting. What is easily seen is that, whenever a portion of a continuum is
removed (or more is added) each point in the material is displaced to a new
position (in the unperturbed original reference frame)~\cite{miofriedman}
\begin{equation}
x^{i\prime}=x^{i}+\xi^{i}\ .
\end{equation}
The new coordinates are obtained by means of a vector displacement field $%
\xi _{i }$. In the continuum a new metric is now induced, which is not the
original Euclidean one $\delta^{ij}$, but
\begin{equation}
g^{ij}=\delta^{ij}+2\varepsilon^{ij} \ ,
\end{equation}
where
\begin{equation}
\varepsilon _{ij}=\frac{1}{2}\left( \frac{\partial \xi _{i}}{\partial x^{j}}%
+ \frac{\partial \xi _{j}}{\partial x^{i}}+\delta ^{lm}\frac{\partial \xi
_{l} }{\partial x^{i}}\frac{\partial \xi _{m}}{\partial x^{j}}\right) \ ,
\label{strain}
\end{equation}
is the (non-linear) strain tensor. It is important to remark that the new
metric, as well as all physical quantities of this description, can equally
well be expressed in terms of the original, undeformed "Lagrangian"
coordinates $x_{i}$, or of the new intrinsic coordinates $x_{i}^{\prime }$,
being the old and the new coordinates numerically identified ~\cite{eshelby}%
. In both cases any point is labelled by the same set of numbers (the
coordinates) plus a vector (the displacement vector at that point) which is
actually zero in the unstrained manifold and non-zero in the strained one.

This framework can be generalized to four dimensions and to spacetime. The
Euclidean basic metric~\footnote{%
Here we have used the standard notation for an $n$-dimensional Euclidean
space with latin indices ranging from $1$ to $n$; on the other hand, for $4$%
-dimensional spacetime, the usual greek indices are used, going from $0$ to $%
3$.} $\delta _{ij}$ is then replaced by the one of Minkowski $\eta _{\mu \nu
}$ and the induced metric is written as~\cite{miofriedman}:
\begin{equation}
g_{\mu \nu }=\eta _{\mu \nu }+2\varepsilon _{\mu \nu }\ .  \label{indotta}
\end{equation}%
Without further details, let us consider an unperturbed (i.e., Euclidean) $4$%
-dimensional space. Then, let us suppose we remove a $4$-sphere
and close the hollow by pulling radially on each point of the
hypersurface of the hole. The situation is described in
Fig.~\ref{fig:defec}. This procedure induces a radial displacement
field represented by a radial four-vector $\xi $. Remarkably,
solving the equations of the elasticity theory with these symmetry
conditions gives, for $\xi $ in four dimensions, precisely a
result like Eq.~(\ref{csidia}): the four-vector $\xi $ has a null
divergence (see for instance Ref.~\cite{eshelby}, Vol. 3 page
107). For spacetime, which implies a Wick rotation in order to
produce the right signature, the induced interval we obtain in
these conditions, once expressed in appropriate coordinates,
corresponds to a typical Robertson-Walker metric
\begin{equation}
ds^{2}=d\tau ^{2}-a^{2}\left( \tau \right) \left[ d\psi ^{2}+\psi ^{2}\left(
d\theta ^{2}+\sin ^{2}\theta d\phi ^{2}\right) \right] \ ,  \label{rob_wal}
\end{equation}%
where $a\left( \tau \right) $ is a non-trivial function and space is flat.
In order to better explain this result, let us start from the general form
of the line element of a Minkowski spacetime expressed in four dimensional
polar coordinates
\begin{equation}
ds^{2}=da^{2}-a^{2}\left[ d\psi ^{2}+\psi ^{2}\left( d\theta ^{2}+\sin
^{2}\theta d\phi ^{2}\right) \right] \ ,
\end{equation}%
where $a$ is now the radial coordinate. The point defect produces, as
written above, a purely radial displacement field, which means that the only
non-vanishing element of the strain tensor (\ref{strain}) is $\varepsilon
_{aa}$. The induced metric, according to (\ref{indotta}), is then

\begin{equation}
ds^{2}=\left( 1+\frac{\partial \xi _{a}}{\partial a}\right) ^{2}da^{2}-a^{2} %
\left[ d\psi ^{2}+\psi ^{2}\left( d\theta ^{2}+\sin ^{2}\theta d\phi
^{2}\right) \right]
\end{equation}
Redefining the radial (actually time) coordinate so that%
\begin{equation}
d\tau =\left( 1+\frac{\partial \xi _{a}}{\partial a}\right) da \ ,
\end{equation}
the old radial coordinate $a$ is expressed as a function of the new time and
the line element becomes (\ref{rob_wal}), as claimed.

In order to clarify the meaning of the $\gamma $ vector, we apply a
procedure typical of the elasticity theory. In a deformed medium the strain
is of course accompanied by a stress. In linear theory, which we now
consider for simplicity, the stress tensor $\sigma _{\mu \nu }$ depends
linearly on the strain tensor $\varepsilon _{\mu \nu }$ (Hooke's law). In
our case the chosen symmetry implies that the radial-radial components of
both tensors (i. e. $\sigma _{aa}$\ and $\varepsilon _{aa}$) be proportional
to each other. The next step is to think of a given solid angle centered at
the singularity, then isolate a portion of it delimited by two transverse
(orthogonal to the radius) (hyper)surfaces. In an equilibrium state the
forces on opposite faces of the boundary of the envisaged piece of material
must be equal in strength. By definition the components of the force on a
small surface (four-dimensional space) are%
\begin{equation}
f^{\mu }=\frac{1}{3!}\sigma ^{\mu \nu }\epsilon _{\nu \alpha \beta \lambda
}dx^{\alpha }\wedge dx^{\beta }\wedge dx^{\lambda }\ .  \label{effe}
\end{equation}%
Calling again in the symmetry, we see that on the \textquotedblleft
bases\textquotedblright\ of our portion of solid angle (\ref{effe}) becomes
\begin{equation}
f^{a}=\mathcal{K}\varepsilon ^{aa}a^{3}d\theta d\phi d\psi \ ,
\end{equation}%
where $\mathcal{K}$ is a constant and $\theta $, $\phi $, $\psi $ are
angles. Now we see that the equilibrium within a given solid angle implies
that $f^{a}$ be independent from $a$ \footnote{%
When considering the example in the text, the forces on opposite sides of
the piece of material must of course be opposite in direction, but we may
consider the force exerted by the external (with respect to the singularity)
medium on a given surface and in this case the direction is everywhere the
same.}; in practice it must be%
\begin{equation}
\varepsilon ^{aa}a^{3}=\text{ constant}\ .  \label{costante}
\end{equation}%
Eq.~(\ref{costante}) is in fact a conservation law. Introducing the
four-vector
\begin{equation}
\gamma =\varepsilon \cdot n\ ,  \label{nuovo}
\end{equation}%
where $n$ is a unit vector orthogonal to a given surface and $\gamma $
represents the flux density of strain. We see that the flux of $\gamma $
across any closed surface is zero, i.e. $\nabla \cdot \gamma =0$. Let us
identify the $\gamma $ in (\ref{nuovo}) with the old one, so its radial
component (the only non-zero component, in our case) will be, as before and
now by virtue of (\ref{costante}) and (\ref{nuovo}),

\begin{equation*}
\chi =\frac{Q^{3}}{a^{3}}\ .
\end{equation*}%
The extended elasticity theory helps us also to find a meaning to
the $Q$ constant. We know that the energy needed to close a void
is given by the product of the pressure times the squeezed volume
$W=pV$. In our case, with the help of the pictorial view of the
situation shown on Fig.~\ref{fig:defec}, we see that the
equivalent of the \textquotedblleft energy\textquotedblright\ is
proportional to
\begin{equation*}
\varepsilon _{aa}\mathcal{A}^{4}\ ,
\end{equation*}%
where $\mathcal{A}$ is the radius of the initial hollow. Calling in (\ref%
{costante}) and (\ref{nuovo}) we see that the "energy" is proportional to $%
Q^{3}\mathcal{A}$. $Q^{3}$ is a measure of the ratio between the work done
to create the defect and its radius.

This is a consistent logical framework. Part of it relies on geometrical
bases, whose meaning is clear in spacetime as well as in three dimensions;
this is the case of the strain tensor and of the definition ~(\ref{nuovo})
for $\gamma$. The rest, i.e. the stress tensor with the related quantities,
is intuitively clear in three dimensions, much less in four, however it is a
tool for arriving to the final interpretation, which remains essentially
geometrical.

%------------------------------------------

\subsection{Equivalent matter distribution}

\label{sbsc:matt_equiv} %------------------------------------------

Once the metric tensor is defined, we can compute from it the Einstein
tensor and, taking the Einstein equations literally, interpret it as being
proportional to the energy-momentum tensor of some matter-energy
distribution responsible for the peculiar metric. Doing this exercise in our
case produces the following effective energy-momentum tensor:
\begin{eqnarray}
T_{\tau }^{\tau } &=&\frac{G_{\tau }^{\tau }}{\kappa }=\frac{3}{\kappa }%
\left( \frac{\dot{a}}{a}\right) ^{2}=\frac{3}{\kappa Q^{2}}\frac{a^{3}}{%
a^{6}+6}e^{1/a^{6}}\ , \\
T_{r}^{r} &=&\frac{G_{r}^{r}}{\kappa }=\frac{2a\overset{\cdot \cdot }{a}+%
\dot{a}^{2}}{\kappa a^{2}}=\frac{6}{\kappa Q^{2}}\frac{(5a^{6}-6)}{%
a^{3}\left( a^{6}+6\right) ^{2}}e^{1/a^{6}}\ , \\
T_{\theta }^{\theta } &=&T_{\phi }^{\phi }=T_{r}^{r}\ ,
\end{eqnarray}%
where it is $\kappa =8\pi G/c^{4}$.

\begin{figure}[t]
\begin{center}
\includegraphics[width=90 mm, height=80 mm]{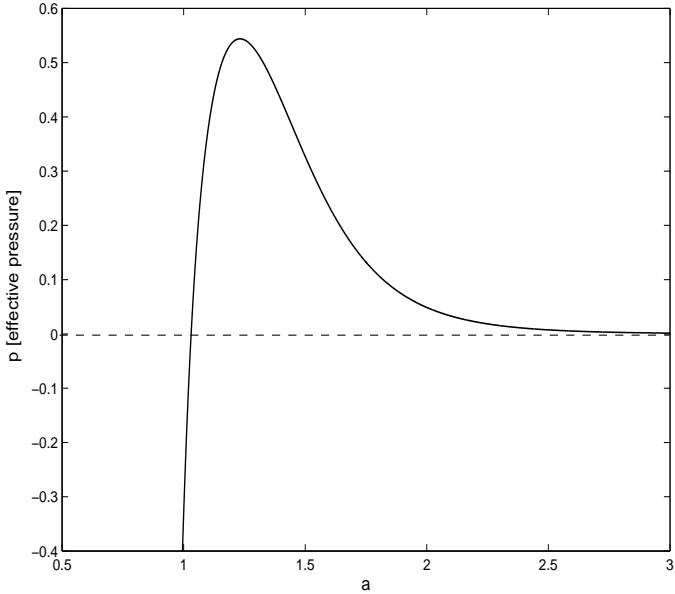}
\caption{\label{fig:pressur}Effective pressure as a function of
the scale parameter $a$. Arbitrary units. The initial negative
values correspond to an inflationary epoch.}
\end{center}
\end{figure}

This energy-momentum tensor has the appearance of the one of a perfect fluid
whose effective matter-energy density is
\begin{equation}
\rho =\frac{3}{\kappa Q^{2}}\frac{a^{3}}{a^{6}+6}e^{1/a^{6}}\ .
\label{eqdens}
\end{equation}%
The corresponding effective pressure is
\begin{equation}
p=\frac{6}{\kappa Q^{2}}\frac{5a^{6}-6}{a^{3}\left( a^{6}+6\right) ^{2}}%
e^{1/a^{6}}\ ,  \label{eqpres}
\end{equation}%
and is represented in Fig.~\ref{fig:pressur}. The initially
negative values
correspond to inflation. %
%%%%%%%%%%%%%%%%%%%%%%%%%%%%%%%%%%%%%%%%%%%%%%%%%%%%%%%%%%%%%%%%%%%%%%%%%
%                   FIG. 4: the effective pressure
%%%%%%%%%%%%%%%%%%%%%%%%%%%%%%%%%%%%%%%%%%%%%%%%%%%%%%%%%%%%%%%%%%%%%%%%%
%%%%%%%%%%%%%%%%%%%%%%%%%%%%%%%%%%%%%%%%%%%%%%%%%%%%%%%%%%%%%%%%%%%%%%%%%%
%
From Eqs.~(\ref{eqdens}) and (\ref{eqpres}) one can immediately obtain the
Equation of State (EOS) of this fluid
\begin{equation}
p=\frac{2(5a^{6}-6)}{a^{6}(a^{6}+6)}\rho \ .  \label{eost}
\end{equation}%
The pressure stays negative up to $a^{6}=6/5$. In the language of dark
energy theories this is equivalent to a peculiar choice of the factor $%
w=p/\rho $. Of course in the case of those theories the equation of state (%
\ref{eost}) would come from a different Lagrangian than ours; however\ here
the comparison is done only at the final stage. What we would like to stress
is that such an equation of state, if sought directly, would appear to be
rather artificial and indeed it is, if thought as pertaining to an actual
"fluid" of any sort.

%-----------------------------

\section{The effect of matter}

\label{sec:matter} %-----------------------------

Let us now verify what the effect of matter is in a spacetime like the one
described before. To this aim, we consider the conceptually simplest
situation and introduce a perfect fluid minimally coupled to the geometry,
so that the total Lagrangian of the problem is (see Ref.~\cite{Schutz} and
references therein)
\begin{equation}  \label{totaction}
S=\int_{\Omega_1}^{\Omega_2} e^{-g_{\mu \nu }\gamma ^{\mu }\gamma ^{\nu
}}Rd\Omega +\kappa \int_{\Omega_1}^{\Omega_2} pd\Omega \ ,
\end{equation}
where $p$ is the pressure of the matter-energy fluid. Following the
traditional approach, we consider that in the present time the fluid is
reduced to an almost incoherent dust, i. e. $p\simeq 0$. In this condition
and with the RW symmetry the matter energy density scales as $\rho =\rho
_{0}/a^{3}$ (matter conservation) so that its contribution to the Lagrangian
is simply a constant: the expansion law is unaffected. When the presence of
the fluid is relevant is in the early epochs where the matter density is
assumed to be negligible with respect to the pressure (radiation dominated
universe). Conservation of entropy together with matter brings about a
pressure that scales as $a^{-4}$%
\begin{equation}  \label{pres}
p=\frac{\psi }{\kappa a^{4}} \ ,
\end{equation}
where $\psi$ is a positive parameter and $\kappa$ has been included for
convenience. From~(\ref{totaction}) and in the case of the RW symmetry, we
obtain the Euler-Lagrange equation:
\begin{equation}
2\left( a+\frac{6}{a^{5}}\right) \ddot{a}+\left( \frac{36}{a^{12}}+1- \frac{%
24}{a^{6}}\right) \dot{a}^{2}-\frac{\psi }{a^{2}} e^{ 1/a^{6} } = 0 \ .
\label{Eq2}
\end{equation}
Condition (\ref{div0}) has again been imposed on the 4-vector $\gamma $, as
before, so that (\ref{csidia}) holds.

Looking for solutions that, in the absence of matter, reduce to the already
known case (\ref{adot}) we pose%
\begin{equation}
\dot{a}=f\left( a\right) e^{\lambda /a^{6}}.  \label{guessadot}
\end{equation}%
Here $f(a)$ is a function of the expansion parameter $a$, and $\lambda $ is
a constant to be determined later. Differentiating Eq.~(\ref{guessadot})
with respect to cosmic time $\tau $ gives
\begin{equation}
\ddot{a}=f\left( f^{\prime }-6f\frac{\lambda }{a^{7}}\right) e^{2\lambda
/a^{6}}\ .  \label{aduedot}
\end{equation}%
where $f^{\prime }\equiv df/da$. %
%%%%%%%%%%%%%%%%%%%%%%%%%%%%%%%%%%%%%%%%%%%%%%%%%%%%%%%%%%%%%%%%%%%%%%%%%
%                   FIG. 5: effect of matter on da/dtau and a(tau)
%%%%%%%%%%%%%%%%%%%%%%%%%%%%%%%%%%%%%%%%%%%%%%%%%%%%%%%%%%%%%%%%%%%%%%%%%
\begin{figure}[t]
\begin{center}
\includegraphics[width=80 mm, height=70 mm]{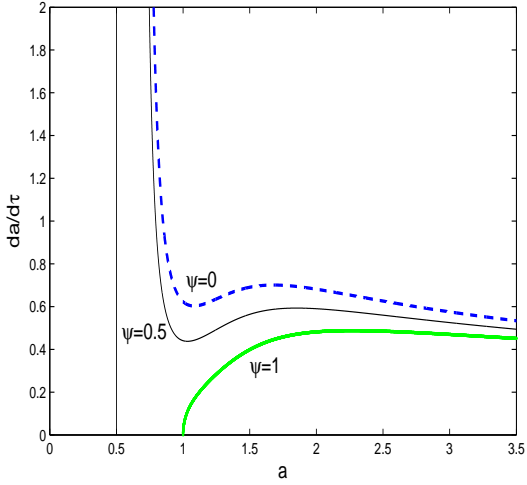}
\caption{\label{fig:confront}Behaviour of the expansion rate of
the universe in the presence of ordinary matter: the solid line
shows the dependence of $a$ on the cosmic time in the case of a
spacetime with matter in subcritical conditions with $\protect\psi
=0.5$. For the sake of comparison we also show the empty spacetime
case ($\protect\psi =0$, dashed line) and a case of supercritical
matter density ($\protect\psi =1$, thick grey line). We note that
the three curves start with an accelerated expansion phase. For
later times this is converted into deceleration, but for
$\protect\psi =0.5$ and $\protect\psi =0$ an effective
\textquotedblleft re-heating\textquotedblright\ occurs.}
\end{center}
\end{figure}
%%%%%%%%%%%%%%%%%%%%%%%%%%%%%%%%%%%%%%%%%%%%%%%%%%%%%%%%%%%%%%%%%%%%%%%%%%
%
Introducing Eqs.~(\ref{aduedot}) and (\ref{guessadot}) into Eq.~(\ref{Eq2})
gives
\begin{align}
& 2\left( a+\frac{6}{a^{5}}\right) ff^{\prime }e^{2\lambda /a^{6}} \\
& +\left( 1+\frac{36-72\lambda }{a^{12}}-\frac{24+12\lambda }{a^{6}}\right)
f^{2}e^{2\lambda /a^{6}}-\frac{\psi }{a^{2}}e^{1/a^{6}}=0\ .  \notag
\end{align}%
Choosing $\lambda =1/2$, this equation becomes
\begin{equation}
2ff^{\prime }\left( a+\frac{6}{a^{5}}\right) +f^{2}\left( 1-\frac{30}{a^{6}}%
\right) -\frac{\psi }{a^{2}}=0\ .  \label{Eqf}
\end{equation}%
The solution of (\ref{Eqf}) is
\begin{equation}
f^{2}=\frac{Aa^{5}-\psi a^{4}}{6+a^{6}}\ ,
\end{equation}%
$A$ is an integration constant.

Finally (\ref{guessadot}) tells us that the expansion rate is:%
\begin{equation}
\dot{a}=a^{2}\left( \frac{a-\psi }{6+a^{6}}\right) ^{1/2}\exp \left( \frac{1%
}{2a^{6}}\right) .  \label{newadot}
\end{equation}
A comparison with (\ref{adot}) fixes $A=1$ and the overall sign of the
formula.

From (\ref{newadot}) we see that the model fails to describe the situation
for%
\begin{equation}
0\leq a< \psi  \label{condi}
\end{equation}%
(imaginary expansion rate). At a smaller scale evidently some more refined
picture is needed.

Looking for the extrema and differentiating (\ref{newadot}) one obtains the
condition:%
\begin{equation}  \label{psi}
a\left( a^{12}-24a^{6}+36\right) +2\psi \left( 9a^{6}-18-a^{12}\right) =0 \ .
\end{equation}

For $\psi =0$ the solutions of (\ref{psi}) are (\ref{estremi}). Studying the
equation for $\psi >0$ we see that three real positive roots exist as far as
$0<\psi <\psi _{c}$, where $\psi _{c}\simeq 0.8$; this means that $\dot{a}$
has three extrema. For $\psi >\psi _{c}$ only one extremum exists. Fig. (%
\pageref{fig:confront}) compares the behaviours of an empty
spacetime with those of one with, respectively, a sub- and a
super-critical matter content.

In practice, when matter is present in the form of a radiation fluid, the
model starts working from a typical value $a=\psi $ of the scale factor.
Initially one has a phase of inflationary accelerated expansion, then the
expansion rate starts decreasing, but after a while, if it is $\psi $
smaller than the critical value, a sort of re-heating happens and the
universe accelerates again its expansion; finally the expansion rate
decreases once more until reaching the $0$ value at infinity. When $\psi
>\psi _{c}$ the initial accelerated expansion is followed by a never ending
deceleration.

The parameter $\psi $ scales as $Q^{2}$, thus, using the estimate of Eq.~(%
\ref{stima}), we see that the critical value, in international units, is
\begin{equation}
\psi _{c}\sim 10^{50}\text{ m}^{2} \ .
\end{equation}
Should we conjecture that the minimal $a_{P}$ value, below which the
classical fluid description fails, is the Planck length, it would be
\begin{equation}
\psi _{P}\sim 10^{-10}\text{ m}^{2} \ .
\end{equation}
Actually one has that (see for instance ~\cite{MTW})
\begin{equation}
\psi =\frac{\kappa }{3}\rho _{r_0}a_{0}^{4} \ ,
\end{equation}
where $\rho _{r_0} $ is the present radiation energy density in the universe
and $a_{0}$ its present scale factor. One usually estimates that $\rho
_{r_0}\sim $ $10^{-13}$ J/m$^{3}$; using for $a_{0}$ the order of magnitude
of $Q$ we obtain
\begin{equation}
\psi \sim 10^{38}\text{ m}^{2} \ ,
\end{equation}
or, in the adimensional form used throughout the paper, $\psi \sim 10^{-10}$%
, well inside the subcritical region. The corresponding minimal scale of the
universe (below which the model is not able to describe what happens) would
be
\begin{equation}
a_{m}\sim 10^{13} \mathrm{m} .
\end{equation}
The conclusion of this section is that the presence of ordinary matter
apparently does not spoil the results obtained in Sec.~\ref{sec:spacetime}
for empty spacetime.

%----------------------------------

\section{The Newtonian limit}

\label{sec:Newton}%----------------------------------

Of course our theory, as any cosmological theory, must prove to be able to
reproduce the known results at the scale of the Solar system and weak
gravitational field, which means that it should possess a Newtonian limit.

In order to prove this it is convenient to start from the action integral (%
\ref{totaction}) and write the general form of the Euler-Lagrange equations
of the theory:%
\begin{align}
& e^{-g_{\alpha \beta }\gamma ^{\alpha }\gamma ^{\beta }}\left(
G_{\mu \nu }-\gamma
_{\mu }\gamma _{\nu }R\right)  \notag \\
& +g_{\mu \nu }e^{-g_{\alpha \beta }\gamma ^{\alpha }\gamma
^{\beta }}\left( 2\gamma ^{\alpha }\gamma _{\alpha ;\sigma }\gamma
_{\lambda }\gamma ^{\lambda ;\sigma }-\gamma ^{\alpha ;\sigma
}\gamma _{\alpha ;\sigma }-\gamma ^{\alpha
}\gamma _{\alpha ;\sigma }^{;\sigma }\right)  \notag \\
& -\frac{1}{2}e^{-g_{\alpha \beta }\gamma ^{\alpha }\gamma ^{\beta
}}g_{\sigma \nu }\left( 2\gamma ^{\alpha }\gamma _{\alpha ;\mu
}\gamma _{\lambda }\gamma ^{\lambda ;\sigma }-\gamma ^{\alpha
;\sigma }\gamma _{\alpha ;\mu }-\gamma
^{\alpha }\gamma _{\alpha ;\mu }{}^{;\sigma }\right)  \label{geneqs} \\
& -\frac{1}{2}e^{-g_{\alpha \beta }\gamma ^{\alpha }\gamma ^{\beta }}g_{\sigma \mu }%
\left[ \mu \rightarrow \nu \right]  \notag \\
& =\kappa \mathsf{T}_{\mu \nu }.  \notag
\end{align}%
The terms on the second and third line are symmetrized in $\mu $ and $\nu $
and $\mathsf{T}_{\mu \nu }$ is the energy-momentum tensor of matter.

Now suppose that%
\begin{equation*}
\mathsf{T}_{\mu \nu }=\mathcal{T}_{\mu \nu }+T_{\mu \nu }
\end{equation*}%
where $\mathcal{T}_{\mu \nu }$ is the energy momentum tensor of the cosmic
fluid and $T_{\mu \nu }$ is the one of a bunch of matter, we assume for
simplicity to have stationary space isotropy around any given point.

Under these conditions we expect the perturbed line element (spacely
isotropic coordinates) to be:%
\begin{equation}
ds^{2}=\left( 1+h_{0}\right) d\tau ^{2}-a^{2}\left( 1+h_{s}\right) \left(
dx^{2}+dy^{2}+dz^{2}\right)  \label{perturb}
\end{equation}%
with $h_{0}$, $h_{s}<<1$ and depending on $r=\sqrt{x^{2}+y^{2}+z^{2}}$ only.

The perturbed metric tensor will perturb the flow lines of the vector $%
\gamma $ field also, inducing the same kind of space symmetry. So we write
for the components of the perturbed vector:%
\begin{eqnarray}
\Upsilon^{0} &=&\chi \left( 1+f_{0}\right)  \label{gammaper} \\
\Upsilon^{i} &=&\chi f_{s}\frac{x^{i}}{r}  \notag
\end{eqnarray}%
where the $f$'s are assumed to depend on $r$ only (as the $h$'s from which
they stem) and $f_{0}$, $f_{s}<<1$ (at least as small as $h$'s).

Recalling that the $\Upsilon$ field has its origin in the cosmic
point defect and that no other defect has been introduced, the
divergencelessness condition $\Upsilon_{;\mu }^{\mu }=0$ must
still
hold. Using the equivalent form $\left( \sqrt{-\mathsf{g}}\Upsilon%
^{\mu }\right) _{,\mu }=0$, the zero order (unperturbed) solution
$\chi =1/a^{3}$, (\ref{perturb}), and the dependences of the $f$'s
and $h$'s, we obtain in the first order approximation
\begin{equation}
\left( f_{s}\frac{x^{i}}{r}\right) _{,i}=0  \label{zero}
\end{equation}%
which implies%
\begin{equation}
f_{s}=\frac{b}{r^{2}}  \label{effesse}
\end{equation}%
$b$ is an integration constant.

A further constraint we can introduce is that the norm of the $%
\Upsilon$ vector remains unchanged. This is because, again, the
vector field depends solely on the existence of a defect and the
global symmetry it
induces. Considering this constraint we write%
\begin{equation*}
\Upsilon_{\mu }\Upsilon^{\mu }=\gamma _{\mu }\gamma ^{\mu }=\chi
^{2}
\end{equation*}%
or, using (\ref{perturb}), (\ref{gammaper}), and stopping at the first order,%
\begin{equation*}
\chi ^{2}=\chi ^{2}\left( 1+2f_{0}+h_{0}\right)
\end{equation*}%
which implies%
\begin{equation}
f_{0}=-\frac{h_{0}}{2}  \label{norma}
\end{equation}

Once these constraints have been implemented we may go back to (\ref{geneqs}%
) and consider the time-time equation:%
\begin{multline*}
e^{-\chi ^{2}}\left( G_{00}-\gamma _{0}^{2}R\right) \\
+g_{00}e^{-\chi ^{2}}\left( 2\gamma ^{\alpha }\gamma _{\alpha ;\sigma
}\gamma _{\lambda }\gamma ^{\lambda ;\sigma }-\gamma ^{\alpha ;\sigma
}\gamma _{\alpha ;\sigma }-\gamma ^{\alpha }\gamma _{\alpha ;\sigma
}^{;\sigma }\right) \\
-e^{-\chi ^{2}}g_{\sigma 0}\left( 2\gamma ^{\alpha }\gamma _{\alpha
;0}\gamma _{\lambda }\gamma ^{\lambda ;\sigma }-\gamma ^{\alpha ;\sigma
}\gamma _{\alpha ;0}-\gamma ^{\alpha }\gamma _{\alpha ;0}{}^{;\sigma }\right)
\\
=\kappa \mathsf{T}_{00}.
\end{multline*}%
The metric tensor is diagonal, which fact implies%
\begin{multline*}
e^{-\chi ^{2}}\left( G_{00}-\gamma _{0}^{2}R\right) \\
+g_{00}e^{-\chi ^{2}}\left( 2\gamma ^{\alpha }\gamma _{\alpha ;i}\gamma
_{\lambda }\gamma ^{\lambda ;i}-\gamma ^{\alpha ;i}\gamma _{\alpha
;i}-\gamma ^{\alpha }\gamma _{\alpha ;i}^{;i}\right) \\
=\kappa \mathsf{T}_{00}.
\end{multline*}%
The next step is to expand everything up to the first order in $h$'s and $f$%
's; drop the zero order terms, which are satisfied by (\ref{csidia}) and (%
\ref{newadot}) with the source $\mathcal{T}_{00}$; use conditions (\ref%
{norma}) and (\ref{zero}). The remaining first order equation is:%
\begin{align}
& -3\frac{\dot{a}^{2}}{a^{2}}h_{0}  \notag \\
& +\frac{1}{a^{2}}\left( h_{s,xx}+h_{s,yy}+h_{s,zz}\right)  \notag \\
& +\frac{\chi ^{2}}{a^{2}}\left( h_{0,xx}+h_{0,yy}+\allowbreak
h_{0,zz}\right)  \label{finale} \\
& +2\frac{\chi ^{2}}{a^{2}}\left(
h_{s,xx}+h_{s,yy}+h_{s,zz}\right) \notag
\\
& =\kappa e^{\chi ^{2}}T_{00}  \notag
\end{align}%
Now, for ordinary time scales the rate of change of $a$, i.e. the Hubble
constant, is extremely small so that we neglect the first term in (\ref%
{finale}). Passing to an orthonormal base (marked by a $\sim $) the factors $%
1/a^{2}$ are absorbed into the space derivatives, so that (\ref{finale})
becomes:%
\begin{equation}
\tilde{\nabla}^{2}h_{s}+\chi ^{2}\left( \tilde{\nabla}^{2}h_{0}+2\tilde{%
\nabla}^{2}h_{s}\right) =\kappa e^{\chi ^{2}}T_{00}  \label{vice}
\end{equation}%
Finally, exploiting the gauge freedom in the choice of the coordinates
(Lorentz gauge with time independent $h$'s), (\ref{vice}) is reduced to%
\begin{equation*}
\tilde{\nabla}^{2}\bar{h}_{0}=-\kappa \frac{e^{\chi ^{2}}}{1+\chi
^{2}}T_{00}
\end{equation*}%
which can be read as the Poisson equation for a Newtonian gravitational
potential with a renormalized coupling constant%
\begin{equation*}
\kappa ^{\ast }=\kappa \frac{e^{\chi ^{2}}}{1+\chi ^{2}}
\end{equation*}%
slowly changing in cosmic times. The vector field $\gamma $ appeares at the
local scale only through its norm $\chi ^{2}$ included in the
renormalization factor of the Newton gravitational constant $G$.

\section{Conclusion and discussion}

\label{sec:conclusions} %----------------------------------

We have applied a heuristic approach to the problem of describing the
behaviour of the universe in its expansion. Instead of introducing new
components in what should correctly be called \textquotedblleft
matter\textquotedblright\ (any scalar or tensor field usually considered is
indeed \textquotedblleft matter\textquotedblright\ in the sense that it
contributes to the right hand side of the Einstein equations and appears
additively in the Lagrangian), we have used a model based on the idea that
the very spacetime is endowed with a property analogous to the internal
viscosity of a fluid. This feature has been treated introducing an
exponential factor in the Langrangian and exploiting from the very beginning
the four-symmetry we think the universe has around the origin. The scalar in
the exponent of the new factor is thus built from a radial (in four
dimensions) vector field. Symmetry considerations suggest that the vector
field can be divergenceless everywhere except in the origin. Solving the
vacuum Einstein equations in these conditions we end up with a global RW
metric with a scale factor depending on cosmic time in such a way to
reproduce an initial inflationary era, followed by a decelerated, then again
accelerated, and finally decelerated expansion of the universe, or, to say
better, of space itself.

Looking for a possible explanation of the \textquotedblleft
friction\textquotedblright\ described by the new vector field, we have had
recourse to a further analogy with ordinary material continua. We have
assimilated empty spacetime to a four-dimensional continuum containing a
pointlike defect, and then we have analyzed the strain and consequent metric
tensor induced by such a defect. The final result is again a RW metric with
a time dependent scale factor. The radial displacement field of this
scenario is divergence-free as the strain flux density is, thus allowing the
identification of the latter with the initial \textquotedblleft
viscous\textquotedblright\ vector field. The model described here presents
results that scale as the\textquotedblleft strength\textquotedblright\ or
\textquotedblleft charge\textquotedblright\ of the center of symmetry, which
may be fixed in order to reproduce the expected age of the universe; this
parameter is geometrically interpreted as representing both the size of the
original void and the \textquotedblleft rigidity\textquotedblright\ of
spacetime, where the final defect (initial singularity) comes from. Besides
this fact, the model, without any further recourse to free parameters,
reproduces reasonably well the observed duration of the acceleration era.
Furthermore, it gives also rise to an initial inflationary era.

One can wonder what changes this theory brings about in the Newtonian limit
and on a local scale. It is easy to see that with respect to this nothing
happens. The relevant quantity in the action~(\ref{action5}) is the scalar $%
\chi ^{2}$ whose rate of change with cosmic time is $2\dot{\chi}/\chi =-6%
\dot{a}/a=-6H$; in practice, with the present value of the Hubble constant,
one has $2\dot{\chi}/\chi \sim -10^{-17}$ s$^{-1}$ whose inverse\textbf{\ }%
correspond\textbf{s} approximately to $3$ billion years. For time intervals
much smaller than the time scale $\mathcal{T\sim }10^{17}$ s the exponential
factor in~(\ref{action5}) is practically constant, thus the known results of
GR hold. If for instance we further introduce in the Lagrangian the typical
space symmetries of the Schwarzschild problem we obtain the corresponding
solution with its Newtonian limit. Only for time intervals comparable with $%
\mathcal{T}$ one can expect changes, which would show up, still using
Schwarzschild as an example, in the form of an adiabatic change in the
unique parameter not fixed by the space symmetry, i.e. the mass of the
source. Many would prefer to state it as a time dependence of the effective
gravitational constant over cosmic times, but the result is the same.
Summing up, we see that GR appears as a short-time approximation of the
theory we propose.

Our final step has been to verify that the addition of ordinary matter in
the form of a fluid (with the densities we obtain from observational data)
does not subvert the behaviour of the universe we obtained for empty
spacetime. Simply, the compound model (spacetime plus matter) starts working
from a minimum scale factor, between the Planck era and the present epoch.

The form initially chosen for the action integral is somehow reminiscent of
other approaches, from string theory to $f\left( R\right) $ theories,
without however coinciding with any of them. We ourselves showed how the
effects may be thought of as being due to an effective fluid with a peculiar
equation of state. As a matter of fact, we obtained our result following
analogies coming from facts of known classical physics and introducing
reasonable (to us) hypotheses, rather than new \textit{ad hoc} entities.

Besides our initial motivation to look for an explanation of the accelerated
cosmic expansion (since our theory is a modification of standard GR), we
obviously would like to verify what the consequences of the new spacetime
Lagrangian are not only for the Newtonian limit discussed above, but also
for such phenomena as the propagation of metric perturbations (gravitational
waves), propagation of electromagnetic waves (modified Einstein-Maxwell
equations), exact solutions in various symmetry conditions (the equivalent
of the Schwarzschild and Kerr solutions) etc. Since ours is a metric theory
having Minkowski both as the tangent and the asymptotic spacetime we do not
expect, at least on not too big scales, relevant changes with respect to the
standard theory. However, the differences could show up both in local high
curvature regions of spacetime and on the large scale behaviour of matter
systems. To explore all these possibilities is our programme for the near
future.

%-------------------------

\section*{Acknowledgments}

\label{sec:ackn} %-------------------------

The authors wish to gratefully acknowledge the help by Alessandro Nagar in
critically reading an early version of the manuscript and contributing to
its reformulation, and by Ninfa Radicella for helping in many formal
calculations.

%--------------
% Bibliography
%-------------

\end{document}